\pdfoutput=1
\documentclass[aip,amsmath,amssymb,jcp,reprint]{revtex4-1}

\usepackage{graphicx}
\usepackage{dcolumn}
\usepackage{bm}

\usepackage[utf8]{inputenc}
\usepackage[T1]{fontenc}
\usepackage{mathptmx}
\usepackage{etoolbox}
\usepackage{subfig}
\usepackage{chemformula}
\usepackage{amsmath}
\usepackage{tikz, pgfplots}
\usepackage{adjustbox}
\usepackage{amssymb}
\usepackage{siunitx}
\usetikzlibrary{positioning}
\usepackage{caption}
\usepackage{amsfonts}
\usepackage{xcolor}
\usepackage[framemethod=tikz]{mdframed}
\usepackage[normalem]{ulem}
\usepackage{lineno}
\newcommand{\overbar}[1]{\mkern 1.5mu\overline{\mkern-1.5mu#1\mkern-1.5mu}\mkern 1.5mu}

\makeatletter
\def\@email#1#2{%
 \endgroup
 \patchcmd{\titleblock@produce}
  {\frontmatter@RRAPformat}
  {\frontmatter@RRAPformat{\produce@RRAP{*#1\href{mailto:#2}{#2}}}\frontmatter@RRAPformat}
  {}{}
}%
\makeatother
\begin{document}

\preprint{AIP/123-QED}

\title{A symmetry-oriented crystal structure prediction method for crystals with rigid bodies}
\author{Qi Zhang}
\affiliation{Department of Physics, Missouri University of Science and Technology, Missouri, Rolla 65401, USA}
\author{Amitava Choudhury}
\affiliation{Department of Chemistry, Missouri University of Science and Technology, Missouri, Rolla 65401, USA}
\author{Aleksandr Chernatynskiy}
\affiliation{Department of Physics, Missouri University of Science and Technology, Missouri, Rolla 65401, USA}
\email{aleksandrc@mst.edu}


\begin{abstract}
We have developed an efficient crystal structure prediction (CSP) method for desired chemical compositions, specifically suited for compounds featuring recurring molecules or rigid bodies. We applied this method to two metal chalcogenides: \ch{Li3PS4} and \ch{Na6Ge2Se6}, treating \ch{PS4} as a tetrahedral rigid body and \ch{Ge2Se6} as an ethane-like dimer rigid body. Initial trials not only identified the experimentally observed structures of these compounds but also uncovered several novel phases, including a new stannite-type \ch{Li3PS4} structure and a potential metastable structure for \ch{Na6Ge2Se6} that exhibits significantly lower energy than the observed phase, as evaluated by density functional theory (DFT) calculations. We compared our results with those obtained using USPEX, a popular CSP package leveraging genetic algorithms. Both methods predicted the same lowest energy structures in both compounds. However, our method demonstrated better performance in predicting metastable structures. The method is implemented with Python code which is available at \url{https://github.com/ColdSnaap/sgrcsp.git}.


\vspace{\baselineskip}
\end{abstract}

\maketitle


\section{introduction}
The crystal structure of a material is arguably the most crucial information, as it directly or indirectly determines all of its properties. With the knowledge of the structure, modern quantum-mechanical methods can reliably calculate those properties prior to compound synthesis. Therefore, crystal structure prediction (CSP) holds immense significance in the field of computational materials design. However, predicting a structure solely based on its chemical composition is highly challenging as it involves classifying a huge number of energy minima on the potential energy surface (PES). Various CSP methods including simulated annealing\cite{kirkpatrick1983optimization,pannetier1990prediction}, minima hopping\cite{amsler2010crystal}, basin hopping\cite{wales1997global}, metadynamics\cite{laio2005assessing}, and random sampling\cite{pickard2011ab} have been developed. All of them suffer from the so-called "curse of dimensionality" due to the usually very high-dimensional PES which is directly related to the number of degrees of freedom for modeling a material system. Hence, the accessible size of the system is still limited and the effectiveness of a CSP method hinges upon its ability to identify and explore the most promising region of PES efficiently. 

To accomplish this objective, some CSP approaches were developed on the basis of a "self-improving" strategy, which assumes that the chemically desired structures often share similar geometric motifs such as bond lengths and the coordination of atoms are located on the low-lying regions of the PES. A typical example of CSP using the self-improving strategy is the genetic algorithm\cite{woodley1999prediction,abraham2006periodic,glass2006uspex,lonie2011xtalopt,trimarchi2007global,yao2008metastable,deaven1995molecular,niesse1996global}, where low-energy structures serve as parents to procreate new candidates by a "crossover" operation. The basic tenet here is that offsprings resemble their parents, and procreation is rewarded for success and an in-depth exploitation is performed in the most promising region on the PES. Another method is the automated assembly of secondary building units (AASBU)\cite{girard200107}. This technique utilizes prior information to enforce connections between individual atoms or rigid bodies, treating them as building units.  By doing so, it effectively reduces the degrees of freedom when performing structure searches which is based on a combination of a simulated annealing procedure and "cost function" minimizations. The success of these approaches demonstrates that the characteristics of recurring motifs in crystal structures play an important role in effectively scanning the promising regions of the PES. This characteristics are commonly found in a class of structures that have attracted considerable attention in various fields: complex metal chalcogenides.

\begin{table*}[t]
\caption{Examples of the WPs in space group $P2_1/c$ (14) and $Pmm2_1$ (31).}
\label{tab:symgroups}
\begin{ruledtabular}
\begin{tabular}{ccccll}
Space & Multiplicity & Wyckoff & Site     & \multicolumn{2}{c}{Coordinates} \\
group &              & letter  & symmetry & &\\
\hline & \\[-1.em]
&4& e & 1 & (1)$x,y,z$ &(2)$-x,-y+1/2,-z-1/2$\\
&&  & & (3)$-x,-y,-z$ &(4) $x,-y+1/2,z+1/2$\\
 $P2_1/c$ &2& d & -1 & (1)$1/2,0,1/2$ &(2)$1/2,1/2,0$\\
(No.14) &2& c & -1 & (1)$0,0,1/2$ &(2)$0,1/2,0$\\
&2& b & -1 & (1)$1/2,0,0$ &(2)$1/2,1/2,1/2$\\
&2& a & -1 & (1)$0,0,0$ &(2)$0,1/2,1/2$\\
\hline & \\[-1.em]
$Pmm2_1$ &4& b & 1 & (1)$x,y,z$ &(2)$-x+1/2,-y,z+1/2$\\
(No.31) &&  & & (3)$x+1/2,-y,z+1/2$ &(4) $-x,y,z$\\
  &2& a & m & (1)$0,y,z$ &(2)$1/2,-y,z+1/2$\\

\end{tabular}
\end{ruledtabular}
\end{table*}

The general formula for complex metal chalcogenides is $ABX$, where $A$ represents a metal (or a combination of 2-3 different metals), $B$ is a main group element, and $X$ denotes a chalcogen. These compounds have emerged as leading candidates for electrolytes in lithium/sodium solid-state batteries\cite{jia2021chalcogenide}, owing to their abundant raw material availability, decent room-temperature ionic conductivity, and lower mechanical stiffness. Complex metal chalcogenides also stand out as some of the most efficient photovoltaics materials\cite{adachi2015earth}, favored for their tunable bandgap, non-toxic and stable nature. In nonlinear optics applications\cite{chung2014metal}, particularly in the IR range where many materials face fundamental limitations, chalcogenides are increasingly utilized. Their application as cathode materials in rechargeable magnesium batteries, seen as a next-generation alternative to lithium-ion batteries due to high energy density, addresses the challenge of slow diffusion kinetics in most cathode materials by offering larger surface areas and shorter migration paths\cite{regulacio2021designing}. Additionally, thermoelectrics\cite{shi2019chalcogenides} and photocatalysis\cite{hausmann2020stannites}, are other possible areas of applications for chalcogenides.
Despite significant interest, much of the chemical space of complex metal chalcogenides remains uncharted, both experimentally and computationally, due to the vast array of possible atomic combinations. This unexplored territory presents substantial opportunities for discovering and developing even more effective materials for a variety of applications.

Many complex chalcogenides are characterized by recurring structural motifs. For instance, the $BX_4$ tetrahedron moiety is extensively studied, with compounds containing this motif being actively developed across various fields\cite{katagiri2008enhanced,adhi2007pulsed,brant2015outstanding,lekse2008synthesis,lekse2009second,navratil2014thermoelectric,sevik2010ab,liu2009improved,pandey2018promising,tsuji2010novel}. Another example is the $B_2X_6$ dimer unit, foundational in recently synthesized non-linear optical materials\cite{balijapelly2022building}. These blocks are notably stable, as demonstrated by one of the synthesis routes called the metathesis reaction. 
Experimentally, this feature of complex chalcogenides enables researchers to synthesize structures with desired properties in a rational manner. Computationally, it allows us to leverage information on bond distances, angles, dihedrals, and connectivity constraints of the rigid blocks within the structure to simplify the structure search.

On the other hand, statistical analysis of modern structural databases on inorganic crystals reveals a striking preference for certain space groups over others. Previous analysis using data from ICSD-2006 indicates that 67\% of all crystal structures are found in only 10\% of all space groups, while 31\% of the space groups are empty or rare \cite{urusov2009frequency}. Furthermore, the preference for occupation of Wyckoff positions (WPs) within space groups has also been summarized \cite{urusov1991crystal}. For example, in most structures of inorganic crystals, anions occupy less symmetric positions than cations. This information can be highly beneficial in locating the most promising regions of the PES for CSP if needed.

Another important question in CSP is how to estimate the stability of candidate structures. First principles methods are commonly used to calculate the energy or enthalpy to determine the thermodynamic stability of candidate structures. However, the cost of local optimizations using first principles methods increases rapidly with the number of atoms in the system. In recent years, machine learning potentials (MLPs) have emerged as a viable alternative for computing various structural properties\cite{schmidt2019recent}. The integration of MLPs in CSP is also gaining attention\cite{tong2020combining} because of their ability to achieve density functional theory (DFT) level accuracy while requiring significantly less computational resources. 

Here, we propose a new crystal structure prediction (CSP) method that enables CSP simulations with rigid bodies and atoms restricted to designated WPs in a specific space group. This method incorporates one of the latest universal MLP: CHGnet \cite{deng2023chgnet}, to enhance its efficiency. It is particularly suitable for metal chalcogenides featuring different types of rigid bodies. We demonstrated its validity and efficiency by applying the proposed method to two systems: \ch{Li3PS4} and \ch{Na6Ge2Se6}. The results were compared to those obtained using USPEX\cite{glass2006uspex}, a popular CSP package that employs genetic algorithm. Both methods identified the same lowest energy structures, with our method identifying more stable metastable candidates within a comparable number of relaxed structures. In Section II, we will detail the algorithms. The application of the method will be introduced in Section III. In the next section we will discuss our results and will conclude the manuscript in Section V.

\section{algorithm}

\begin{table*}[t]
\caption{\label{tab:Li3PS4cases}
Numbers of combinations for \ch{Li3PS4} in different space groups with a maximum of three \ch{PS4} rigid bodies in the unit cell.}
\begin{ruledtabular}
\begin{tabular}{cccccc}
Space group & Combinations & Space group & Combinations & Space group & Combinations\\
\hline
\colorbox{blue!30}{1} & 3 & 2 & 127 & 3 & $>$1000\\
\colorbox{blue!30}{4} & 1 & \colorbox{blue!30}{5} & 12 & 6 & 256\\
\colorbox{blue!30}{7} & 1 & \colorbox{blue!30}{8} & 2 & 10 & $>$1000\\
\colorbox{blue!30}{11} & 20 & 13 & 76 & 16 & $>$1000\\
17 & 96 & \colorbox{blue!30}{18} & 12 & 21 & 88 \\
23 & 76 & 25 & $>$1000 & \colorbox{blue!30}{26} & 12 \\
27 & 96 & 28 & 39 & \colorbox{blue!30}{30} & 12 \\
\colorbox{blue!30}{31} & 2 & \colorbox{blue!30}{32} & 12 & \colorbox{blue!30}{34} & 12 \\ 
\colorbox{blue!30}{35} & 20 & \colorbox{blue!30}{38} & 20 & \colorbox{blue!30}{44} & 16\\
47 & $>$1000 & 51 & 124 & 59 & 24\\
\colorbox{blue!30}{75} & 20 & 77 & 39 & 81 & >1000\\
82 & 28 & 84 & 60 & \colorbox{blue!30}{85} & 16\\
\colorbox{blue!30}{86} & 8 & 89 & 206 & \colorbox{blue!30}{90} & 16\\ 
93 & 330 & \colorbox{blue!30}{94} & 8 & 99 & 28\\
\colorbox{blue!30}{100} & 6 & \colorbox{blue!30}{101} & 16 & \colorbox{blue!30}{102} & 3\\
105 & 48 & 111 & $>$1000 & 112 & 270\\
113 & 22 & \colorbox{blue!30}{114} & 4 & 115 & $>$1000\\
116 & 64 & 117 & 52 & 118 & 52\\
119 & 28 & \colorbox{blue!30}{121} & 6 & 125 & 26\\
\colorbox{blue!30}{129} & 16 & 131 & 90 & 132 & 38\\
\colorbox{blue!30}{134} & 10 & \colorbox{blue!30}{137} & 4 & 143 & $>$1000\\
\colorbox{blue!30}{144} & 1 & \colorbox{blue!30}{145} & 1 & \colorbox{blue!30}{146} & 2\\
147 & 34 & 149 & $>$1000 & 150 & 130\\
\colorbox{blue!30}{151} & 12 & \colorbox{blue!30}{152} & 12 & \colorbox{blue!30}{153} & 12\\
\colorbox{blue!30}{154} & 12 & 156 & $>$1000 & 157 & 64\\
158 & 33 & \colorbox{blue!30}{159} & 10 & \colorbox{blue!30}{160} & 2\\
168 & 23 & \colorbox{blue!30}{171} & 12 & \colorbox{blue!30}{172} & 12\\
\colorbox{blue!30}{173} & 10 & 174 & $>$1000 & 177 & 130\\
180 & 76 & 181 & 76 & 183 & 27\\
\colorbox{blue!30}{185} & 3 & \colorbox{blue!30}{186} & 10 & 187 & $>$1000\\
189 & 206 & 195 & 23 & \colorbox{blue!30}{197} & 1\\
\colorbox{blue!30}{201} & 1 & \colorbox{blue!30}{208} & 3 & \colorbox{blue!30}{215} & 17\\
\colorbox{blue!30}{217} & 1 & \colorbox{blue!30}{218} & 3 & \colorbox{blue!30}{224} & 1
\end{tabular}
\end{ruledtabular}
\begin{minipage}{18cm}
\vspace{0.1cm}
\begin{flushleft}
\footnotesize Note: We have applied the proposed CSP method to the highlighted space groups, which have 20 or fewer combinations. Combinations exceeding 1000 are marked as '$>1000$'. Space groups not included in the table have no viable cases.
\end{flushleft}
\end{minipage}
\end{table*}

\subsection{Symmetry of rigid bodies}

Determining the ground state structure in CSP represents a global optimization challenge within “phase space” of all possible structures, characterized by a dimensionality of $3N-6$, where $N$ signifies the number of atoms in the system. The subtraction of 6 accounts for the structure's translational and rotational degrees of freedom. As the number of atoms in the crystal unit cell increases, this dimensionality rises linearly. Consequently, the volume of the phase space increases exponentially, a phenomenon often referred to as the "curse of dimensionality." However, the inherent symmetry within the crystal structure offers an advantage, allowing for a reduction in the number of independent atomic coordinates and, consequently, the dimensionality of the phase space wherein the structure resides. This dimensionality reduction directly translates to fewer exploratory attempts required to find the minimum energy configuration, thereby making identifying the ground state structure more efficient\cite{pickard2011ab,lyakhov2013new}.

For a crystal with a given symmetry, the atomic positions are classified by WPs. The most general WP has the highest multiplicity, and the atoms found on it do not lie on any symmetry elements. The remaining sites are the so-called special positions, and the atoms that occupy these sites reside on symmetry elements of the cell. A WP is unique if all its coordinates are fixed. meaning they do not contain variables. A unique position can host only a single atom, whereas a non-unique WP can accommodate multiple atoms. For instance, consider space group $P2_1/c$ (14), The coordinates for its WPs are delineated in table \ref{tab:symgroups}. Here, the most general WP is $4e$, which has a multiplicity of 4 and is not unique. In contrast, other WPs ($2a$, $2b$, $2c$, and $2d$) are unique, each with multiplicity of 2. However, the most crucial aspect of WPs for our applications is the site symmetry. This is because an arbitrary object can occupy specific WP in periodic crystals only if the symmetry group of that object contains the site symmetries as a subgroup.

In crystallography, atoms are typically considered as spherically symmetric point particles. As a result, the full rotation group of any atom inherently includes the site symmetry of a WP as a subgroup. This allows atoms to be located anywhere within the unit cell. However, for molecules or rigid bodies, their own symmetry groups impose restrictions on WP compatibility, determining if a rigid block can be situated at a specific WP. Additionally, the alignment between the symmetry group of the rigid block and the site symmetries frequently limits the block's orientation, effectively reducing the dimensionality of the phase space.

Consider a perfect tetrahedron rigid block within a crystal structure of space group $Pmm2_1$(31). A perfect tetrahedron possesses symmetries of 3-fold rotation, 2-fold rotation, mirror plane and 4-fold rotoinversion. We reference the site's symmetries for space group $Pmm2_1$(31) in table \ref{tab:symgroups}. The general position $4b$ has only the identity operation, which is a subgroup of any group, thus allowing the tetrahedron to occupy this WP. However, with a multiplicity of four, the crystal must contain four tetrahedra in its unit cell. The identity operation does not restrict the orientation of the tetrahedron, allowing it to rotate freely around three general axes. Besides the general WP, this group also includes special WP $2a$, which contains a mirror plane. The combination of mirror and identity operations forms a subgroup of the tetrahedral symmetry group, making this WP compatible with the tetrahedron. With a multiplicity of two, two tetrahedra must be present in the unit cell. The mirror plane's specific orientation necessitates that the tetrahedron's mirror plane coincides with it, limiting its rotation to a single axis perpendicular to the mirror plane. Since neither WPs are unique, multiple tetrahedra can be placed at each of those positions. In practice, the total number of atoms or rigid bodies in the unit cell is capped due to limited computational resources, limiting possible combinations. If we restrict the crystal to contain only two tetrahedra per unit cell, the sole option is placing them at $2a$ WP, with only three independent variables describing their positions: y and z coordinates of the WP, and the rotation angle in the mirror plane. For four tetrahedra, two configurations are possible: i) all four at the general position $4b$ with arbitrary orientation or ii) two sets of tetrahedra at WP $2a$. Both cases require six independent coordinates to specify the unit cell. On the other hand, space group $P2_1/c$ (14) has special WPs whose symmetry elements consist solely of inversion, as shown in table \ref{tab:symgroups}. This operation is not a symmetry of a tetrahedron. Therefore, the only option is to place all tetrahedra at general positions when searching for compounds with tetrahedra in this space group. This sets the minimum number of tetrahedra at four in the unit cell. Consequently, space group 14 will be disregarded if we limit the total number of tetrahedra to two. Analyzing the symmetry of the rigid body and determining its compatibility WPs in a space group is the first step of the simulation as shown in figure \ref{fig:flowchat}(a). This process is implemented using PyXtal \cite{fredericks2021pyxtal}, which is similar to the Wyckoff Alignment of Molecules (WAM) method \cite{darby2020ab} proposed recently to restrict the searching area on the PES.

\subsection{Enumerating possible configurations}

With the idea of the symmetry, it is feasible to enumerate all possible atom arrangements for a certain compound composition as in figure \ref{fig:flowchat}(b). For example, let us list all possible combinations of atomic arrangements to WPs for \ch{Li3PS4}, where \ch{PS4} is a tetrahedron rigid body, in the space group $Pmm2_1$(31). From the previous discussion, there must be an even number of the tetrahedral units in the primitive cell, and for this example we will consider the smallest number of the stoichiometric units, i.e. two. In general, the size of the unit cell is limited by the computing power available for local minimization through DFT. In practice, we set the maximum number of atoms in the cell, and progressively consider possible number of stoichiometric units in the cell up to this limit. \ch{PS4} rigid bodies can be at both $2a$ and $4b$ WPs based on site symmetries. However, since $4b$ WP has multiplicity of four, and we are considering only 2 stoichiometric units per cell we are left with the only possibility of \ch{PS4} units at $2a$. The number of tetrahedral units sets the number of lithium atoms in the unit cell to six. Taking into account the multiplicities of the WPs, these atoms can be distributed only in the following two combinations: \{$4b$,$2a$\} and \{$2a$,$2a$,$2a$\}. We list the numbers of possible combinations for every space group in Table \ref{tab:Li3PS4cases} with the restriction that the crystal contains not more than three tetrahedra per primitive cell. For all the missing space groups, it is not possible to arrange up to three stoichiometric units of \ch{Li3PS4} in that space group.

While the symmetry restrictions from the rigid body and the crystal space group reduce the dimensionality of the phase space for the global optimization, the number of the possible arrangements of rigid block and atoms among WPs is likely to be very large in some space groups. Examining the Table \ref{tab:Li3PS4cases} we see that \ch{Li3PS4} in the space group $Pmma$(51) has 124 possible arrangements of the rigid blocks and lithium atoms. Considering the structure prediction method involves many structure relaxations through DFT for each arrangement. It renders the search within this space group very time consuming. We circumvent this difficulty by taking the subgroups of a space group into consideration.

A subgroup of a particular space group contains only some of the symmetry operations of that space group. It means that there are fewer restrictions for the atomic positions from the symmetry considerations. Effectively, setting the structure within the subgroup will replace (large) enumerated number of possible combinations for atomic positions with Monte-Carlo search, but in higher-dimensional phase space. If the global minimum structure possesses the symmetry of the original group, then this will appear as “accidental” symmetry in the best structure if the algorithm is successful. In the example of the tetrahedron rigid block above, when we are trying to find the lowest energy structure in the space group $Pmma$(51), we can look at its subgroup $P2_1$(4) instead, which only has 1 possible arrangements of atoms/rigid blocks among WPs. Of course, that process can be continued all the way to the space group $P1$ which contains only the identity operation. This will be equivalent to ignoring the symmetries of the crystal and corresponding reduction in the dimensionality of the phase space.

\subsection{Generation of the initial structure}
The initial structure for global minimization is crucial for the efficient convergence of the optimization process. Besides atomic positions, the structure is also defined by crystal basis vectors. For convenience, it is more practical to use three lattice constants and three 
angles rather than basis vectors, and the space group symmetry imposes additional constraints on these parameters. A classic example is a cubic lattice where $a=b=c$ and $\alpha=\beta=\gamma=90.0^\circ$, leaving only one unique parameter to define. The initial values of lattice 
constants and angles are set randomly within the constraints that the structure should adhere to a specified density and desired lattice system. 

At this stage, the allocation of atoms and rigid blocks to the WPs is already established. Therefore, we generate initial structures by randomly assigning atomic coordinates to all WPs and randomly orienting the rigid blocks as in figure \ref{fig:flowchat}(e), if permissible. 

\subsection{Global optimization}

\begin{figure*}[t]%
    \centering
    \includegraphics[width=18cm]{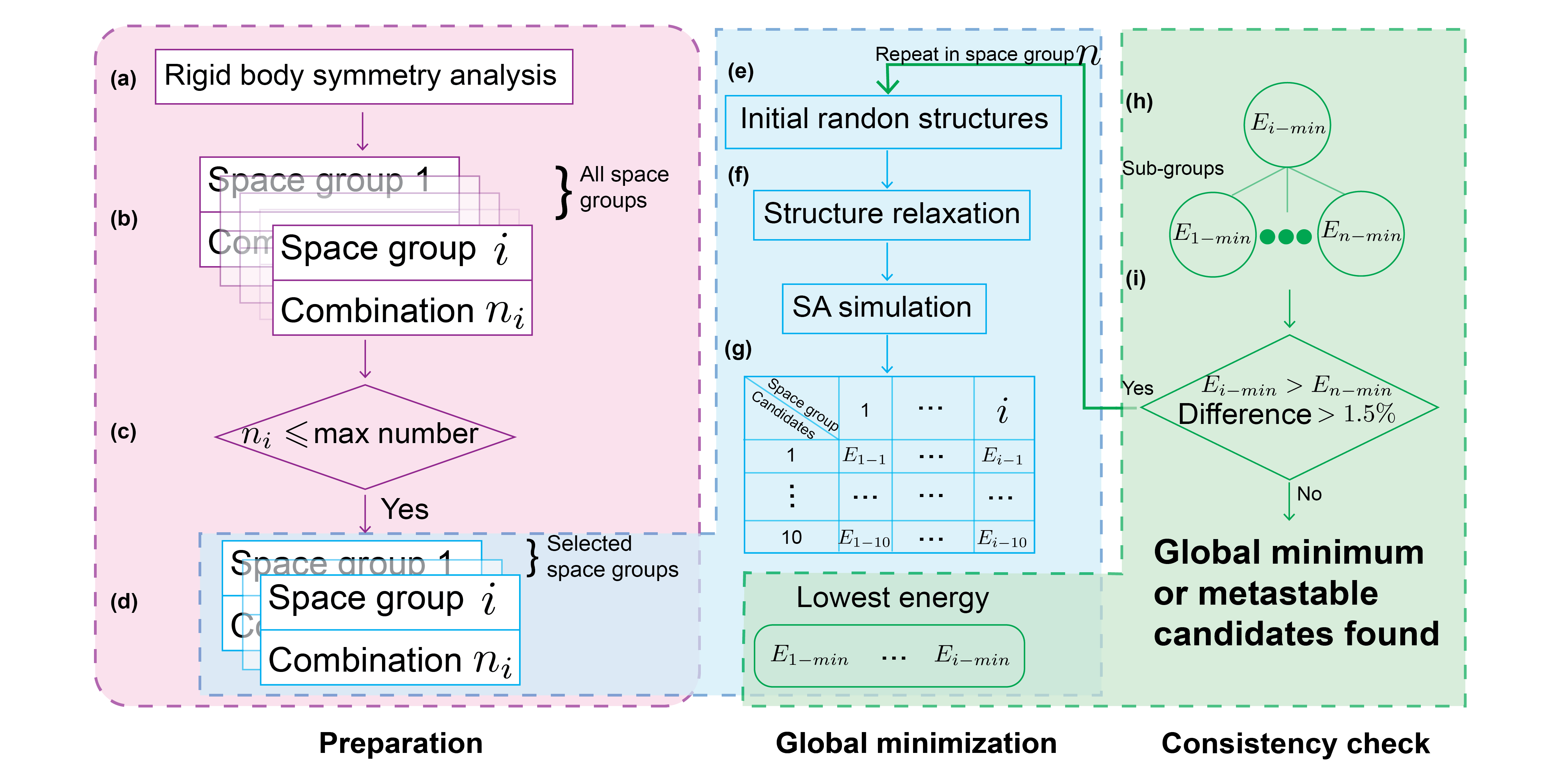}
    \caption{\label{fig:flowchat} Schematic illustration of the simulation procedure. (a) Analysis of rigid body symmetry to determine suitable WPs for the rigid body's placement. (b) Enumeration of potential combinations within each space group. (c) Exclusion of space groups that exceed the predetermined maximum number of combinations. (d) Selection of specific space groups for conducting global minimization. (e) Generation of initial random structures within the selected space groups. (f) Preliminary structure relaxation using DFT to refine the rigid body geometry and cell shape. (g) Storage of the ten lowest energy structures for each atomic arrangement in every selected space group. (h) Energy comparison between different groups and their respective subgroups. (i) Implementation of criteria to ascertain the identification of the global minimum or potential metastable candidates.}
\end{figure*}

\begin{figure*}[t]%
    \includegraphics[width=18cm]{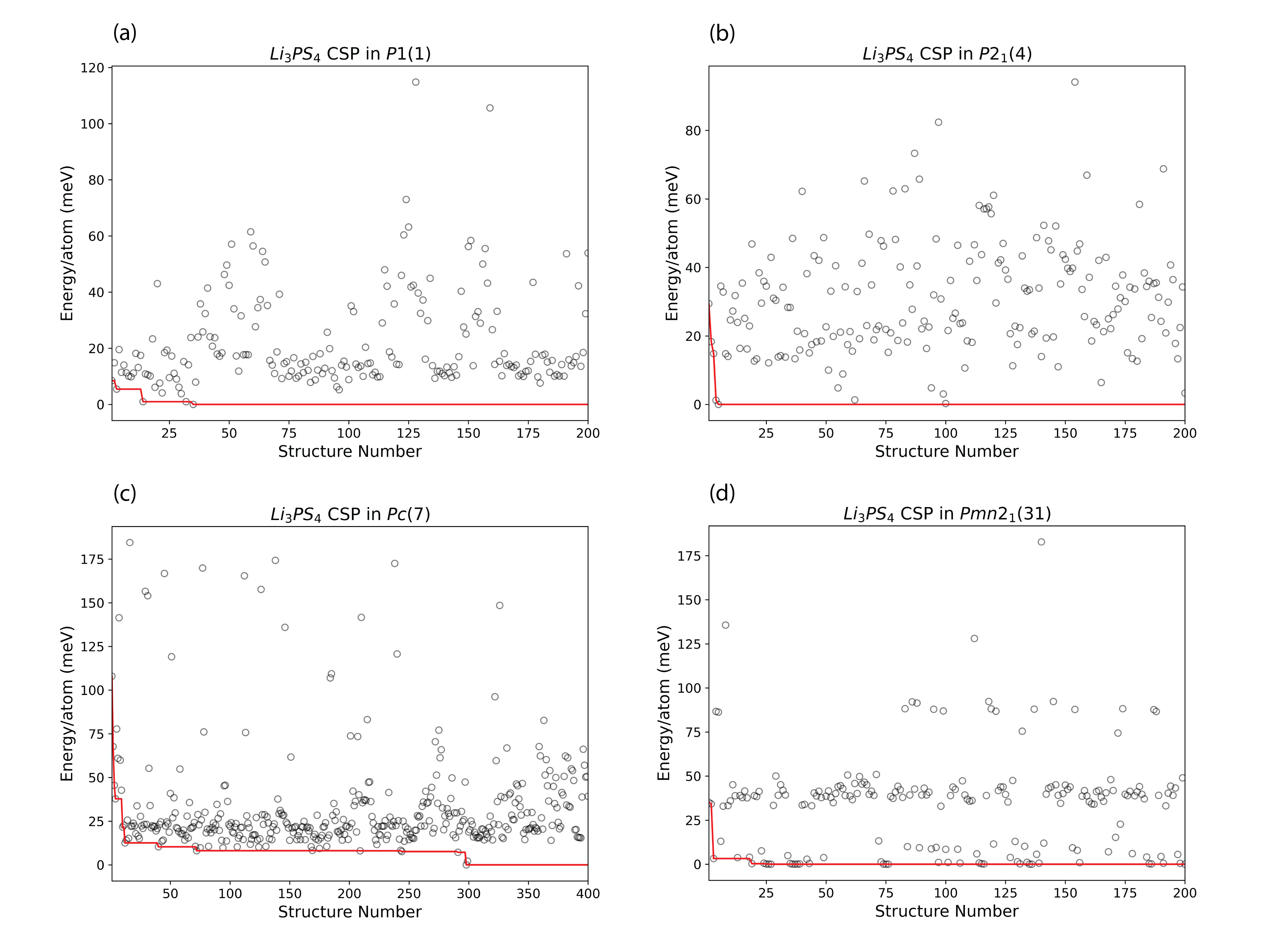}
    \caption{\label{fig:ps4_csp}The relaxed structure numbers and their corresponding energies above the ground state for \ch{Li3PS4} are plotted for four different space groups:(a) $P1$(1), (b) $P2_1$(4), (c) $Pc$(7) and (d) $Pmn2_1$(31).}
\end{figure*}

\begin{figure*}[t]%
    \includegraphics[width=18cm]{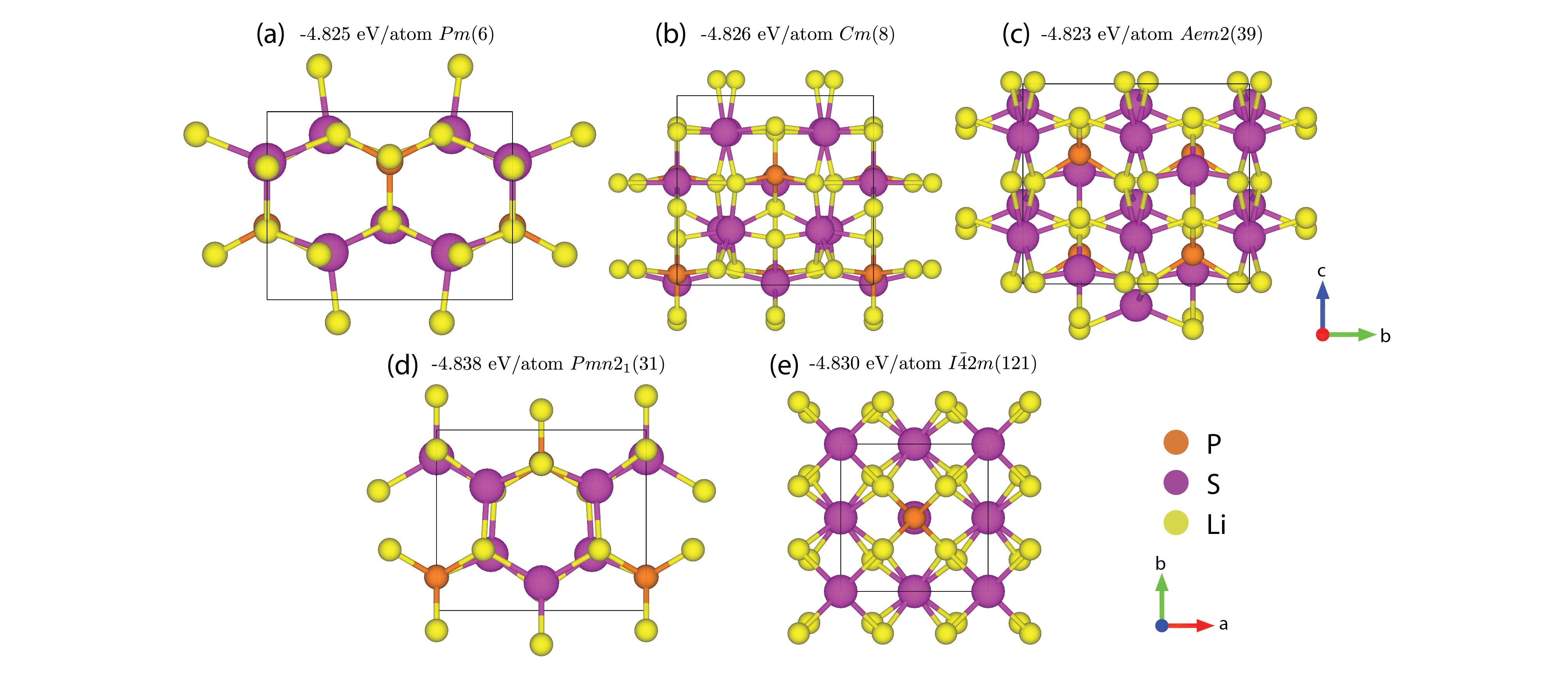}
    \caption{\label{fig:ps4_struc}Low-energy structures identified for \ch{Li3PS4}. Structures (a), (b), (d), and (e) are newly predicted, with (d) being the observed structure.}
\end{figure*}

\begin{figure}[h]%
    \includegraphics[width=8cm]{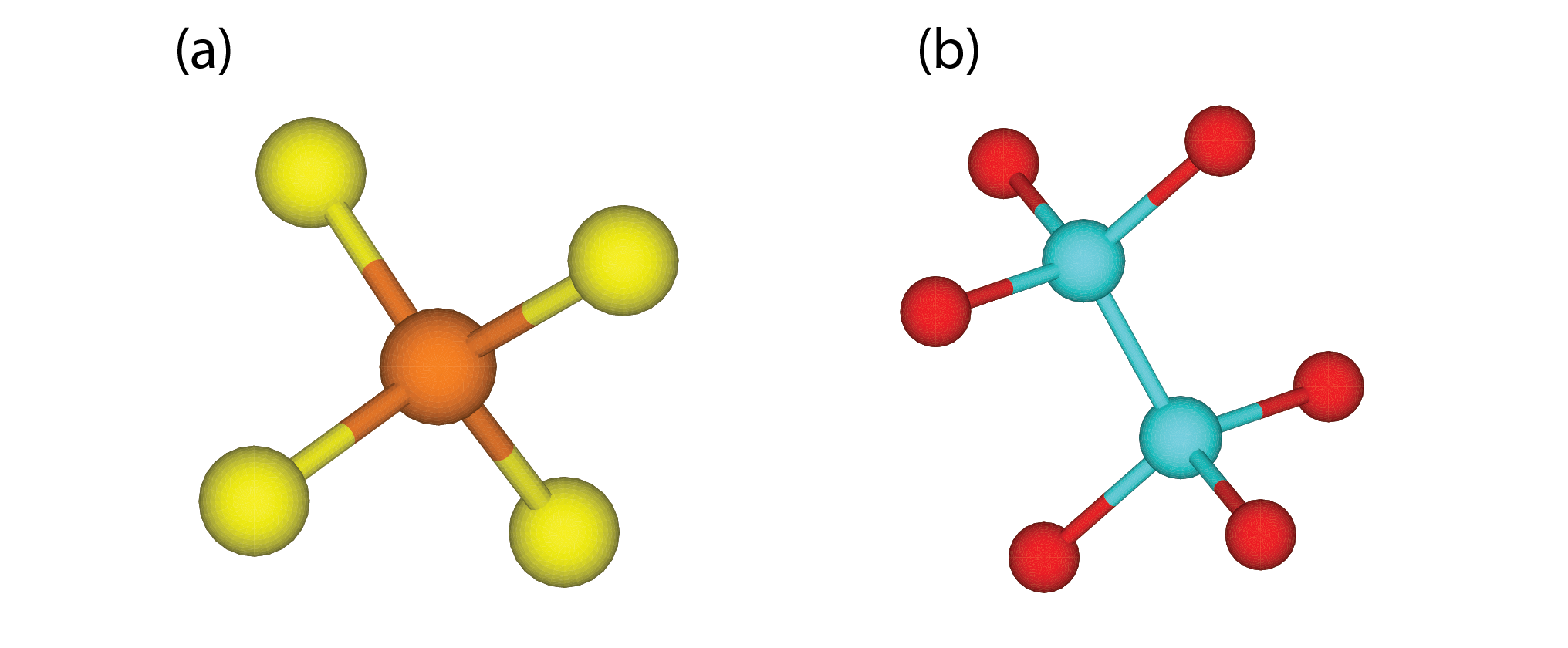}
    \caption{ \label{fig:RigidBlocks}(a) Tetrahedral rigid body \ch{PS4} (b) Ethane-like dimer rigid body \ch{Ge2Se6}}
\end{figure}

\begin{figure*}[t]%
    \includegraphics[width=18cm]{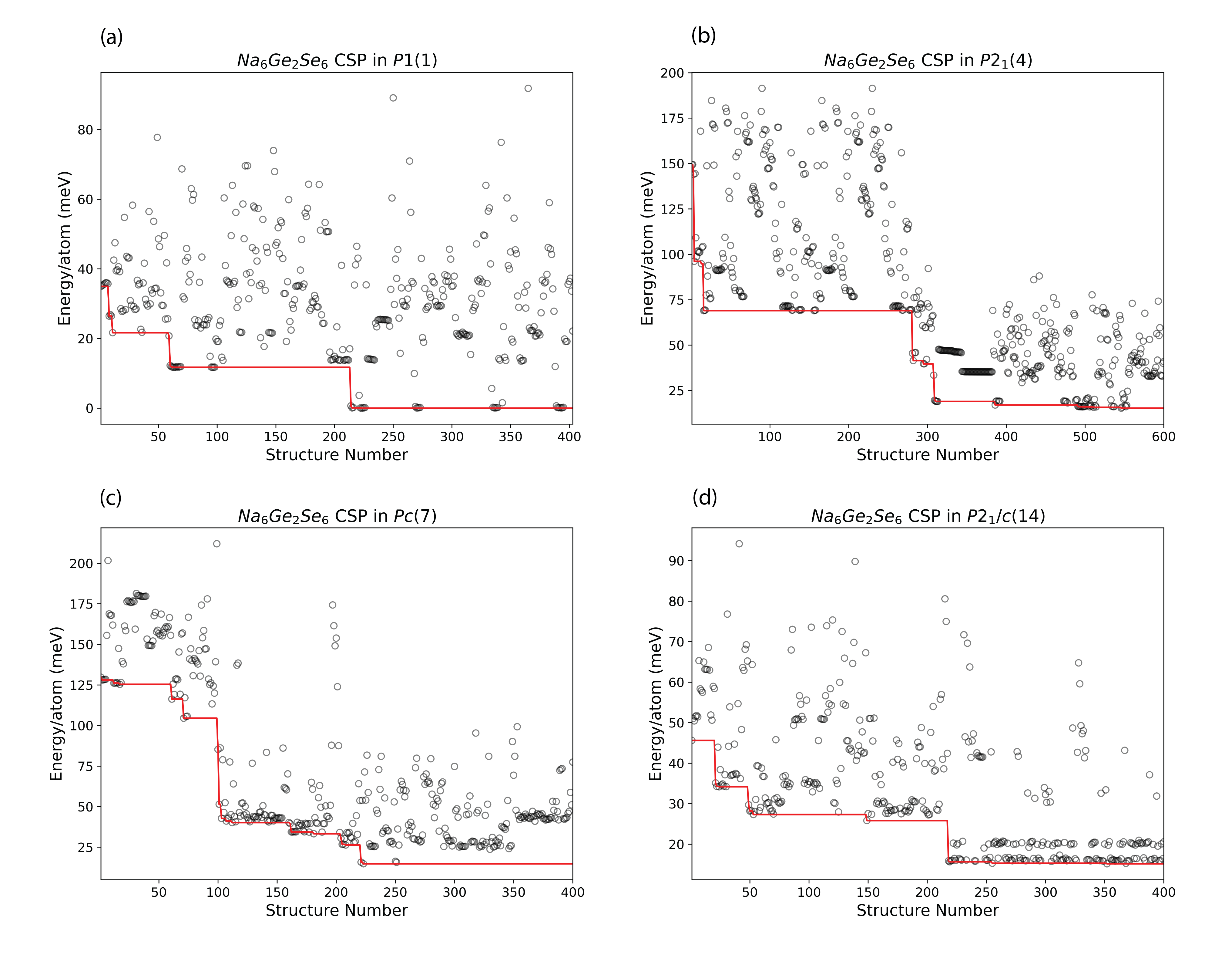}
    \caption{\label{fig:gese_csp}The relaxed structure numbers and their corresponding energies above the ground state of \ch{Na6Ge2Se6} are plotted for four different space groups:(a) $P1$(1), (b) $P2_1$(4), (c) $Pc$(7) and (d) $P2_1/c$(14).}
\end{figure*}

\begin{figure*}[t]%
    \includegraphics[width=18cm]{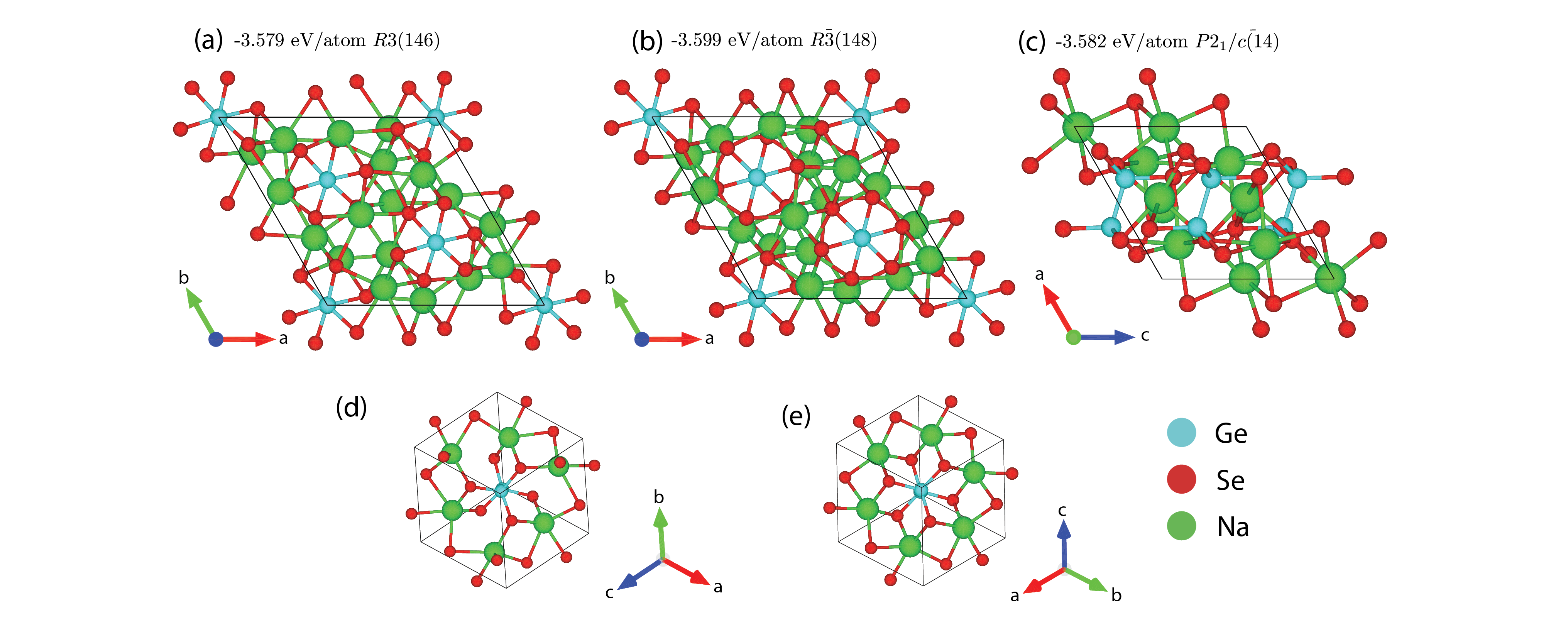}
    \caption{\label{fig:gese_struc}Low-energy structures identified for \ch{Na6Ge2Se6}. Structures (d) and (e) are the primitive cells for (a) and (b) respectively. Structure (c) is the observed phase.}
\end{figure*}

Once we specify the configuration of atoms and rigid bodies among the Wyckoff positions, we carry out a global optimization of the energy via a modified simulated annealing (SA) method.

The SA closely mirrors the natural process where a molten system's energy progressively reduces as the temperature decreases, ultimately arriving at the global 
minimum configuration - the most stable crystalline phase. The SA is initiated by simulating the system's dynamic evolution at some initial temperature using 
either molecular dynamics or a Monte-Carlo scheme. The initial temperature has to be chosen so that there are sufficient kinetic energy to overcome energy barriers 
between local minima. The temperature is then progressively lowered as the system continues its evolution. The temperature lowering gradually restricts 
transitions between the minima that have barriers greater than the current temperature, and eventually the system arrives and stays in the global minimum, if 
certain conditions are met\cite{geman1984stochastic}. In this work we employ a Monte-Carlo scheme to simulate the dynamics, which contains three ingredients: The probability 
distribution $g(\Delta x_i,T)$ to select a new random trial structure, the probability distribution $p(T,\Delta E)$ to accept the new structure and the 
cooling schedule $T(k)$. Here, $\Delta x_i$ is a change of the parameter $x_i$ to be optimized, i.e. symmetry-restricted positions of the atoms and rigid 
bodies, $T(k)$ is a temperature $T$ at the step $k$ of the simulation, $\Delta E$ is the energy difference between the new and the current structures.

If one chooses Boltzmann SA, i.e. for each parameter the function 
$g(\Delta x_i,T)$ is
\begin{equation}
g(\Delta x_i,T)=(2\pi{T^{-\frac{1}{2}}})exp[-\Delta{x_i^2/(2T^2)}]
\end{equation}
and the acceptance probability as 
\begin{equation}
p(T,\Delta E)=\frac{1}{1+exp({\Delta}E/T)}
\end{equation}
Then it can be shown\cite{geman1984stochastic}, that logarithmic reduction of the temperature, i.e. $T(k)={T_0}/{\ln{k}}$ guarantees the convergence of the method. In practice, 
however, logarithmic reduction is too slow, i.e. it requires a prohibitive number of simulation steps, especially if the energy is evaluated via the first principles calculations. Thus, the convergence of the method is not assured. We note, that there are other choices possible for the distributions $g(\Delta x_i,T)$ and $p(T,\Delta E)$ which permit faster cooling 
schedule\cite{kirkpatrick1983optimization,szu1987fast}, however applicability of these approaches for our optimization problem is beyond the scope of this work.

To improve the chances of finding the global minimum we introduce the following modifications to the SA method. SA methods typically record simulation results at predetermined intervals, followed by a local energy minimization, with the final post-minimization structures serving as candidate structures for the global minimum. Our approach extends beyond merely storing these candidates. We leverage the configurations of these candidates, including lattice parameters and atomic positions, as the starting point for the subsequent SA round. This strategy capitalizes on the fact that the first principles software, VASP\cite{kresse1996efficiency,kresse1996efficient}, that we use in this work, conveniently implements symmetry-restricted structure optimization. Consequently, we can update the lattice parameters post-minimization instead of constantly perturbing the lattice randomly, allowing for a further reduction in dimensionality. During the SA process, only the atoms undergo random movement within their designated WPs, while the six lattice parameters remain constant.

Additionally, the bond lengths within the rigid bodies are adjusted based on the DFT optimization. This adjustment is particularly advantageous when only the general geometry of the rigid body is known, and an approximate bond length is set prior to the simulation. However, it is possible for the rigid body to disintegrate following a minimization step. To address this, we implement a rigid body integrity check after each energy minimization. If the rigid body is found to have broken apart, we store the relaxed structure in case the structure with broken rigid bodies results in a lower energy phase. However, the subsequent simulation will continue with the last relaxed structure that has intact rigid bodies.

Secondly, the term $\Delta E$ in the acceptance probability equation exhibits varying sensitivities to the optimization parameters. It is thus advantageous to scale $\Delta x_i$ by the inverse of the energy change due to the variation of the parameter $x_i$ alone. These sensitivities are periodically evaluated throughout the simulation run.
Lastly, our findings suggest that gradually cooling the system is beneficial. However, since our methodology does not depend on a specific cooling schedule to ensure convergence, the exact nature of this dependence is not critically important. We implement a cooling schedule as $T(k) = T_0 \times 0.99^{k/N}$, where $N$ represents the number of steps between subsequent local optimizations. In typical simulation runs for the applications discussed below, we use the following parameter values: the initial temperature is set to $T_0 = 2$ K, the total number of steps is $k_{\text{max}} = 10^4$, and local optimization occurs every $N = 50$ steps. 

After finishing the simulations for all space groups with feasible distributions of atoms and rigid blocks among WPs, we store the ten lowest energy structures for each combination, as depicted in figure \ref{fig:flowchat}(g), for potential metastable structures. Subsequently, we conduct a consistency check in chosen space groups that exhibit low-energy structures, using the lowest energy structure from other space groups to determine whether the global minimum or metastable candidates have been identified. This step is also illustrated in figure \ref{fig:flowchat}.
For a specific space group, simulations within its subgroups should yield energies equal to or lower than those obtained within the group itself, provided the optimization in that subgroup was successful. Thus, we compare the lowest energies found in available subgroups of a given group. If they are consistent and the lowest among all discovered energies, it strongly suggests that the ground state has been identified. However, if energies within subgroups are higher than in the parent group by more than 1.5\% from our experience, this indicates incomplete optimization in the subgroup, prompting further calculations with new random initial structures. The number of loops run for this process can be determined as needed; in this work, we set it to three. If, after three attempts, the energy within a subgroup remains lower than that within the group, the optimization success remains uncertain.
Such a consistency check is vital because, in our implementation, SA does not guarantee the discovery of the global minimum, as previously discussed.


\subsection{Implementation details}
An open-source toolpack is available at \url{https://github.com/ColdSnaap/sgrcsp.git} to automate the entire CSP process. The input files consist of three parts: an input.txt file containing user-defined parameters, rigid body/molecule files, and first principles calculation inputs. While the full documentation is currently being prepared, simple instructions and examples are available on the public GitHub repository to help users reproduce the results presented in this paper.

The entire code is implemented using Python 3. Structure-related properties such as symmetry analysis, are calculated with pymatgen\cite{ong2013python} and ASE\cite{larsen2017atomic}. Structure generations are carried out with PyXtal\cite{fredericks2021pyxtal}. To support scientific computing and data processing tasks, it also requires NumPy\cite{oliphant2006guide}, SciPy\cite{virtanen2020scipy} and Spglib\cite{togo2018texttt}.

The CSP process involves two types of structure energy evaluations. The first type is the calculation of energy for unrelaxed structures, which is required for the structure update procedure of SA. In addition to first principles methods, we offer the option of using the MLP CHGnet\cite{deng2023chgnet} for faster energy calculations. The second type is the energy calculation with structure relaxation. For this part, we prioritize accuracy and therefore use the first principles method implemented with the Vienna Ab initio Simulation Package (VASP) \cite{kresse1996efficiency,kresse1996efficient} (version 5.4.4).

The reported results are a combination of using VASP and CHGnet for the SA process. For calculations performed with VASP which employs Projector-Augmented-Wave method for the treatment of the core electrons\cite{blochl1994projector,kresse1999g}. The Perdew-Burke-Ernzerhof (PBE)\cite{perdew1996generalized} functional was used. Integration over the Brillouin zone was carried out using the tetrahedron method with Blöchl corrections (ISMEAR=-5)\cite{blochl1994improved}. For structure relaxation, a $3\times3\times3$ K-point mesh and plane wave basis sets with an energy cutoff of 520 eV were employed. The convergence threshold for the electronic self-consistent loop was set at \num{2e-5} eV. For single point calculations, the K-point mesh was reduced to $2\times2\times2$, with an energy cutoff of 420 eV, and the convergence criteria were loosened to \num{2e-4} eV.


\section{applications}

To test our method, we carried out the global optimization of \ch{Li3PS4} and \ch{Na6Ge2Se6}, wherein \ch{PS4} and \ch{Ge2Se6} form a tetrahedral and an ethane-like dimer rigid block respectively, as illustrated in figure \ref{fig:RigidBlocks}. \ch{Li3PS4} is a thoroughly studied system, with three distinct phases experimentally identified\cite{homma2011crystal} and subsequently verified via computational studies\cite{kam2023crystal}. Regrettably, two of these phases, the $\alpha$- and $\beta$-phases, exhibit fractional occupancy of the crystallographic sites, rendering them unidentifiable through our methodology, thus we expect to find only a ground state $\gamma$-phase.
\ch{Na6Ge2Se6} has gained lesser attention in the research community and only a single phase has been experimentally observed\cite{eisenmann1985oligoselenidogermanate}. As we describe below, our method has not only successfully found the $\gamma$-phase \ch{Li3PS4} and experimentally observed \ch{Na6Ge2Se6} phase, but also yielded several candidates for metastable structures. Notably, we identified several structures for \ch{Na6Ge2Se6} that possesses lower energy than that of the observed structure.




Our analysis starts with the \ch{Li3PS4} system, featuring a rigid block that is the previously mentioned tetrahedron composed of phosphorus and sulfur atoms. As its symmetries have been thoroughly addressed in the methodology section, we will not delve into them here. Instead, for a detailed account of potential combinations for arranging up to three stoichiometric units in various space groups, we refer the reader to table \ref{tab:Li3PS4cases}. Space groups not included in the table are those where allocating the \ch{PS4} tetrahedron and lithium atoms among the WPs is not viable. Some groups, like 10, 16, 25, and 47, present an exceptionally high number of combinations. To maintain practicality, we set a cap of 20 as the maximum number of combinations for our calculations, thus omitting space groups that surpass this limit and focusing instead on their subgroups.

Table \ref{tab:Li3ps4Energy} enumerates the lowest energy values found within each space group setting, while Figure \ref{fig:ps4_struc} illustrates five low-energy structures along with their corresponding space groups. The lowest energy recorded across all space groups is -4.838 eV/atom, observed in space groups $Pmn2_1$(31), $Pc$(7), $P2_1$(4), and $P1$(1). These structures are identical to one another, representing the $\gamma$-phase \ch{Li3PS4} illustrated in Figure \ref{fig:ps4_struc}(d). From Figure \ref{fig:ps4_csp}, it can be observed that the simulation identified the $\gamma$-phase in the first 200 relaxed structures (10k steps of SA) in $Pmn2_1$(31), $P2_1$(4) and $P1$(1), with the exception of $Pc$(7). Consequently, further simulations are necessary for $Pc$(7) to ensure it visits the possible lowest energy phase ($\gamma$-phase). In the second trial with a new initial random structure, the simulation successfully identified the $\gamma$-phase \ch{Li3PS4}, with no structure of lower energy found after another 200 relaxed structures. According to our consistency check criteria, the fact that space group $Pmn2_1$(31) found the same structure as its subgroups ($Pc$(7), $P2_1$(4) and $P1$(1)) adds another layer of confidence that the $\gamma$-phase is the lowest energy configuration in $Pmn2_1$(31) and possibly the ground state of this chemical composition.

\begin{figure*}[t]%
    \includegraphics[width=18cm]{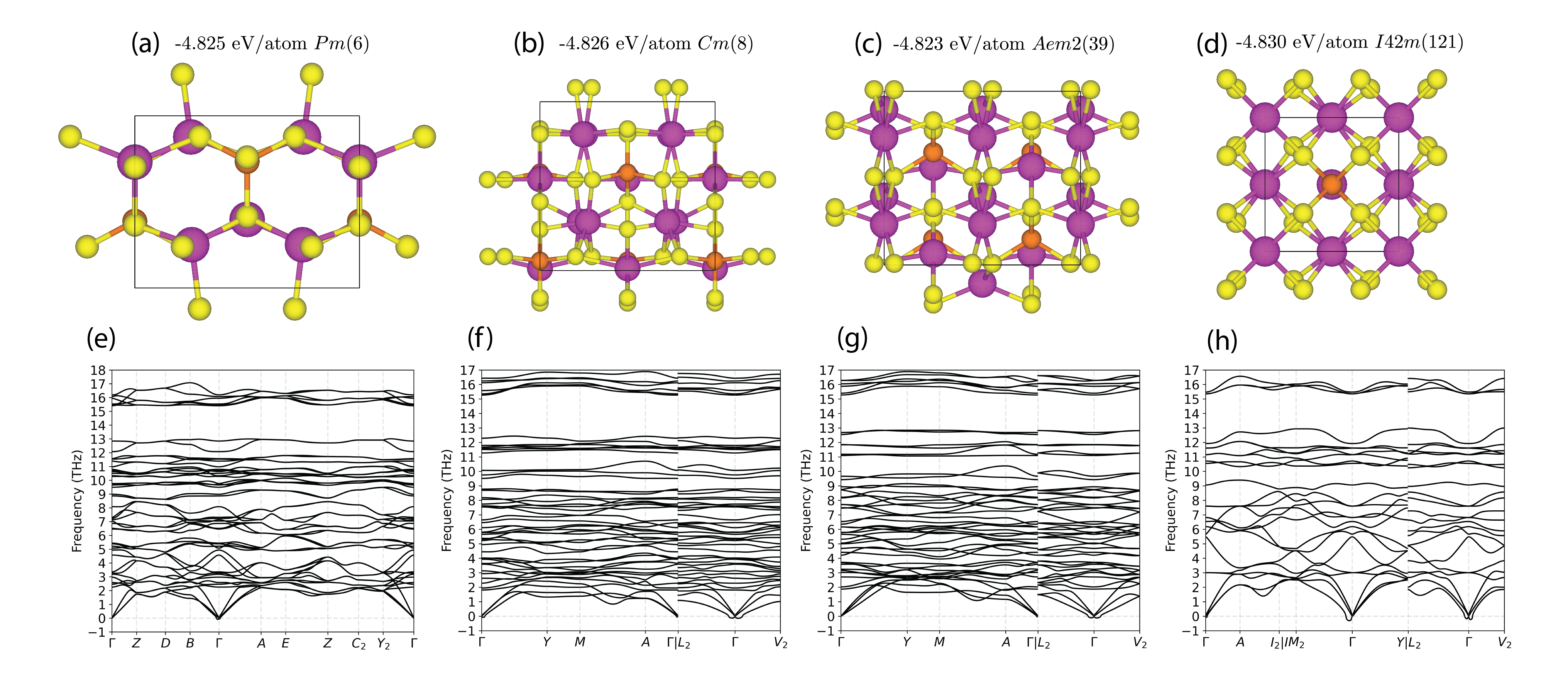}
    \caption{\label{fig:ps_phonon}The phonon band structure of the predicted \ch{Li3PS4} structures. (e) is for structure (a), (f) is for structure (b), (g) is for structure (c) and (h) is for structure (d)}
\end{figure*}

In addition to the $\gamma$-phase, we identified several metastable candidates as illustrated in Figure \ref{fig:ps4_struc}(a), (b), (c) and (e). The next lowest energy structure is found 8 meV/atom above the $\gamma$-phase and crystallizes in space group $I\overbar{4}2m$(121). This structure represents the well-known stannite structure, common in the $M_xBX_4$ compound family including systems like \ch{Cu2ZnSnS4}\cite{ito2015copper,wang2014device,katagiri2001development}, \ch{Zn2CuGaS4}\cite{bindi2020richardsite}, \ch{Cu2HgSnSe_{x}Te4_{-x}}\cite{navratil2014thermoelectric}, and \ch{Cu2FeSnS4}\cite{hausmann2020stannites}. The subsequent lowest energy structure is found in $Cm$(8), 12 meV/atom above the $\gamma$-phase, followed by structures in $Pm$(6) and $Aem2$(39), with 13 meV/atom and 15 meV/atom above the $\gamma$-phase respectively. The phonon band structures for these predicted structures are shown in Figure \ref{fig:ps_phonon} to demonstrate their dynamical stability. However, it is possible for a low-energy structure to be dynamically unstable. For instance, the lowest energy structure found in space group $P2_1/m$(11) is unstable, even though it is only 12 meV/atom above the $\gamma$-phase.

We now turn our attention to \ch{Na6Ge2Se6}. This compound features an ethane-like, dumbbell-shaped rigid block composed of germanium and selenium atoms as depicted in Figure \ref{fig:RigidBlocks}. The symmetry elements of this rigid body include mirror planes, 2-fold and 3-fold rotation axes, and inversion symmetry. The stoichiometric unit of \ch{Na6Ge2Se6} comprises 14 atoms, exceeding the 8 atoms in \ch{Li3PS4}, leading us to limit the unit cell size to two stoichiometric units for this compound. Details of the possible combinations are shown in Table \ref{tab:GeSecases}, and Table \ref{tab:GeSeEnergy} presents the lowest energies identified for each space group where calculations were conducted.

\begin{figure*}[t]%
    \includegraphics[width=18cm]{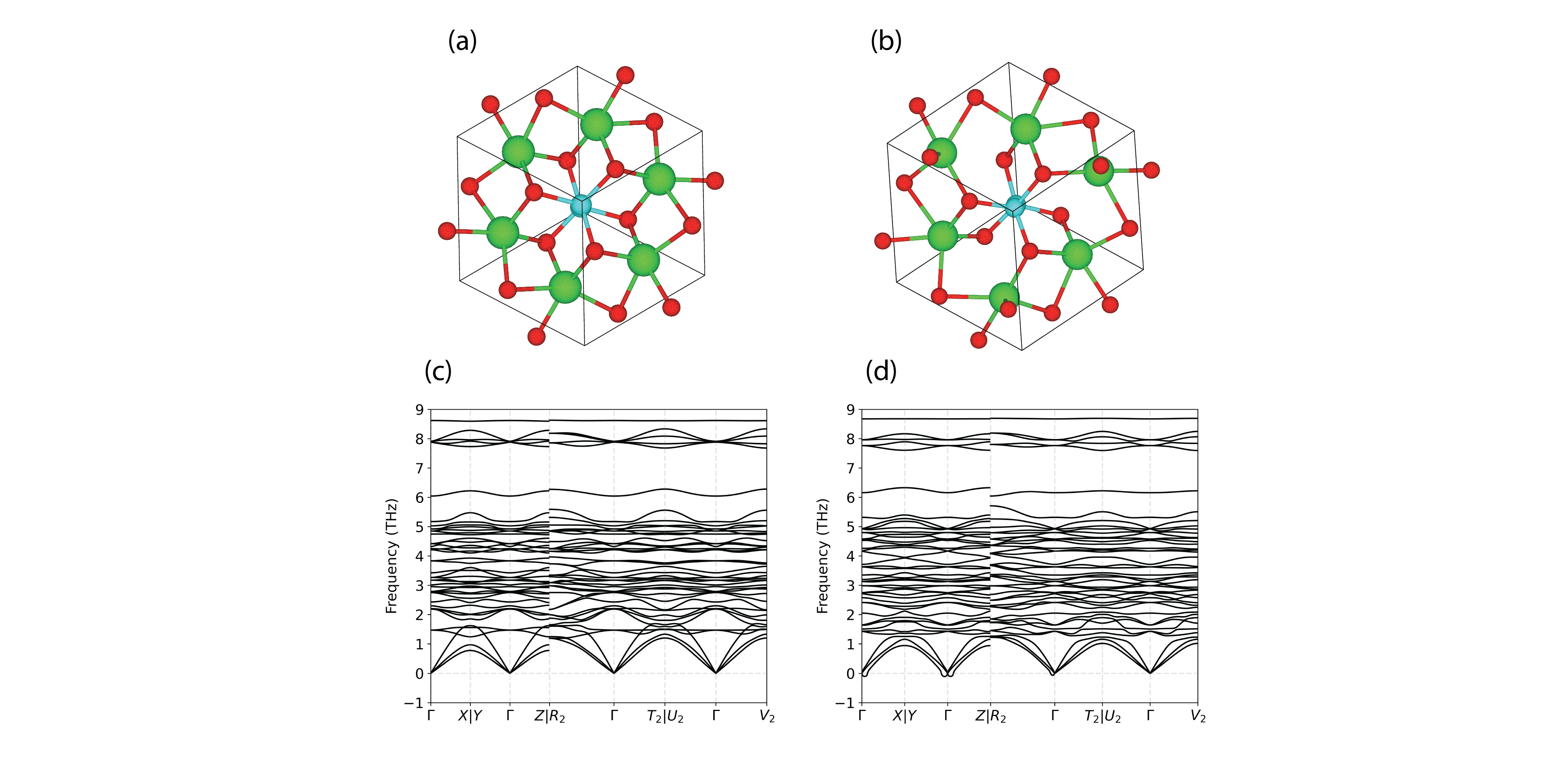}
    \caption{\label{fig:gese_phonon}The phonon band structures of the predicted \ch{Na6Ge2Se6} phases are shown. (c) corresponds to structure (a), which is the lowest energy phase we identified and is illustrated in Figure \ref{fig:gese_struc}(a). (d) corresponds to structure (b), which is the metastable candidate illustrated in Figure \ref{fig:gese_csp}(b).}
\end{figure*}

The CSP results for space groups $P1$(1), $P2_1$(4), $Pc$(7) and $P2_1/c$(14) are presented in Figure \ref{fig:gese_csp}. The lowest energy structures identified in groups $P2_1$(4), $Pc$(7), and $P2_1/c$(14) are identical and it is illustrated in Figure \ref{fig:gese_struc}(c).
This \ch{Na6Ge2Se6} phase was experimentally observed in 1985\cite{eisenmann1985oligoselenidogermanate} and has an energy of -3.582 eV/atom. Notably, the CSP results for space group $P1$(1) indicates the presence of several structures with significantly lower energy than the observed phase. Most of these structures crystallize in space group $P\overbar{1}$(2) following symmetry analysis (Figure \ref{fig:gese_sym2}). The lowest energy phase found from $P1$(1) crystallizes in space group $R\overbar{3}$(148) as illustrated in Figure \ref{fig:gese_struc}(b), with an energy 17 meV/atom lower than the observed phase. Additionally, another structure crystallizes in space group $R3$(146) as shown in Figure \ref{fig:gese_struc}(a), is a metastable phase candidate with an energy 3 meV/atom higher than the observed phase. The primitive cells of these two phases are illustrated in Figures \ref{fig:gese_struc}(d) and \ref{fig:gese_struc}(e). The calculation of their phonon band structures, presented in Figure \ref{fig:gese_phonon}, indicates their dynamical stability. The lowest energy phase in $R\overbar{3}$(148) differs markedly from the observed one, suggesting that the experimentally observed structure may be metastable. To ensure this finding is not a simulation artifact, we conducted additional relaxations with VASP using the PBE0 hybrid functional \cite{perdew1996generalized, adamo1999toward}, which provides a higher level of accuracy. The results indicate that its energy is 15 meV/atom lower than that of the observed phase, reaffirming the validity of our findings.


\section{discussion}

\begin{figure}[h]%
    \includegraphics[width=8cm]{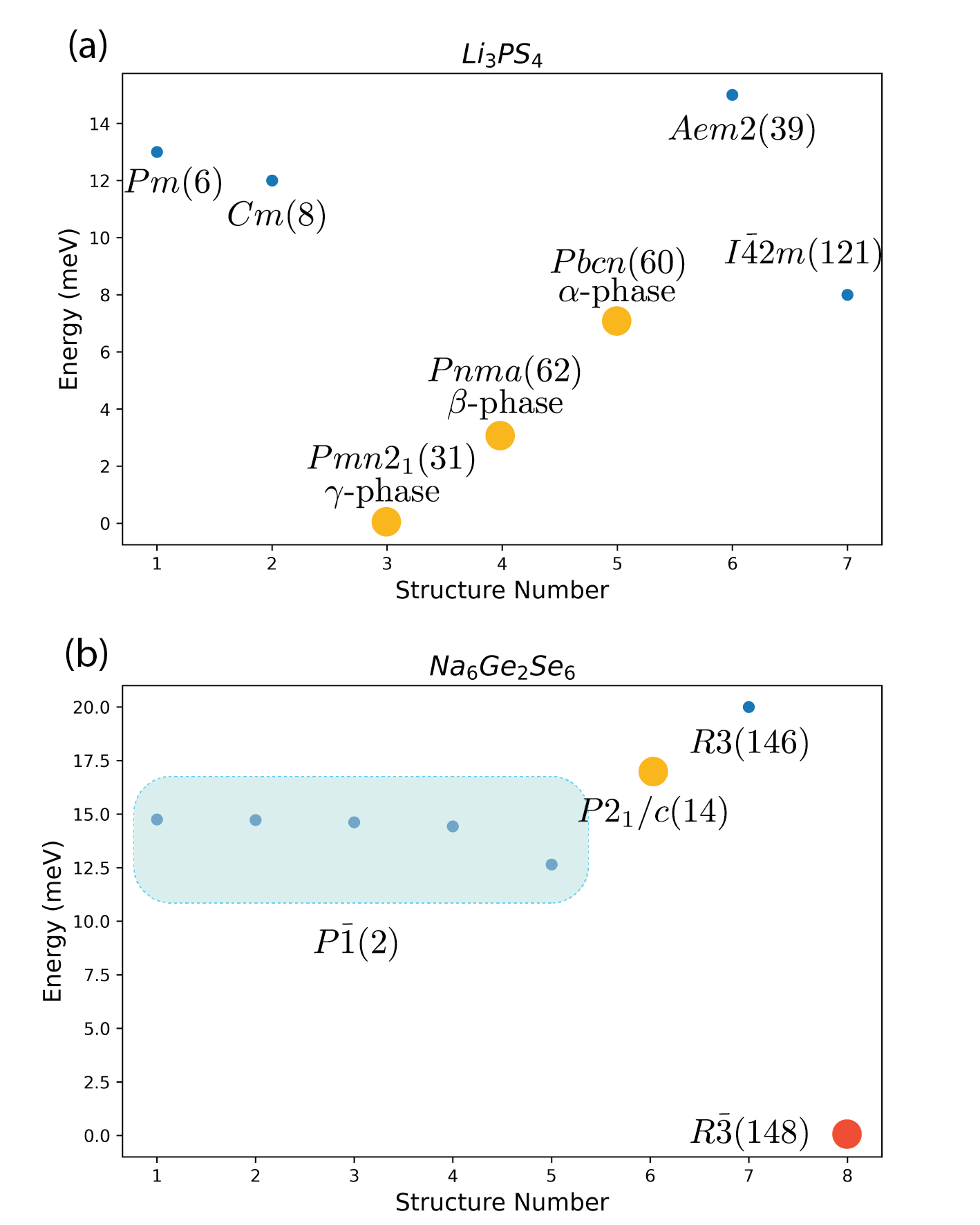}
    \caption{ \label{fig:energy_compare}Energy comparison between predicted and observed phases for (a) \ch{Li3PS4} and (b) \ch{Na6Ge2Se6}. The orange dots in (a) and (b) represent the experimentally observed phases, while the red dot in (b) indicates the predicted phase with lower energy than the observed phase.}
\end{figure}

\begin{figure}[h]%
    \includegraphics[width=8cm]{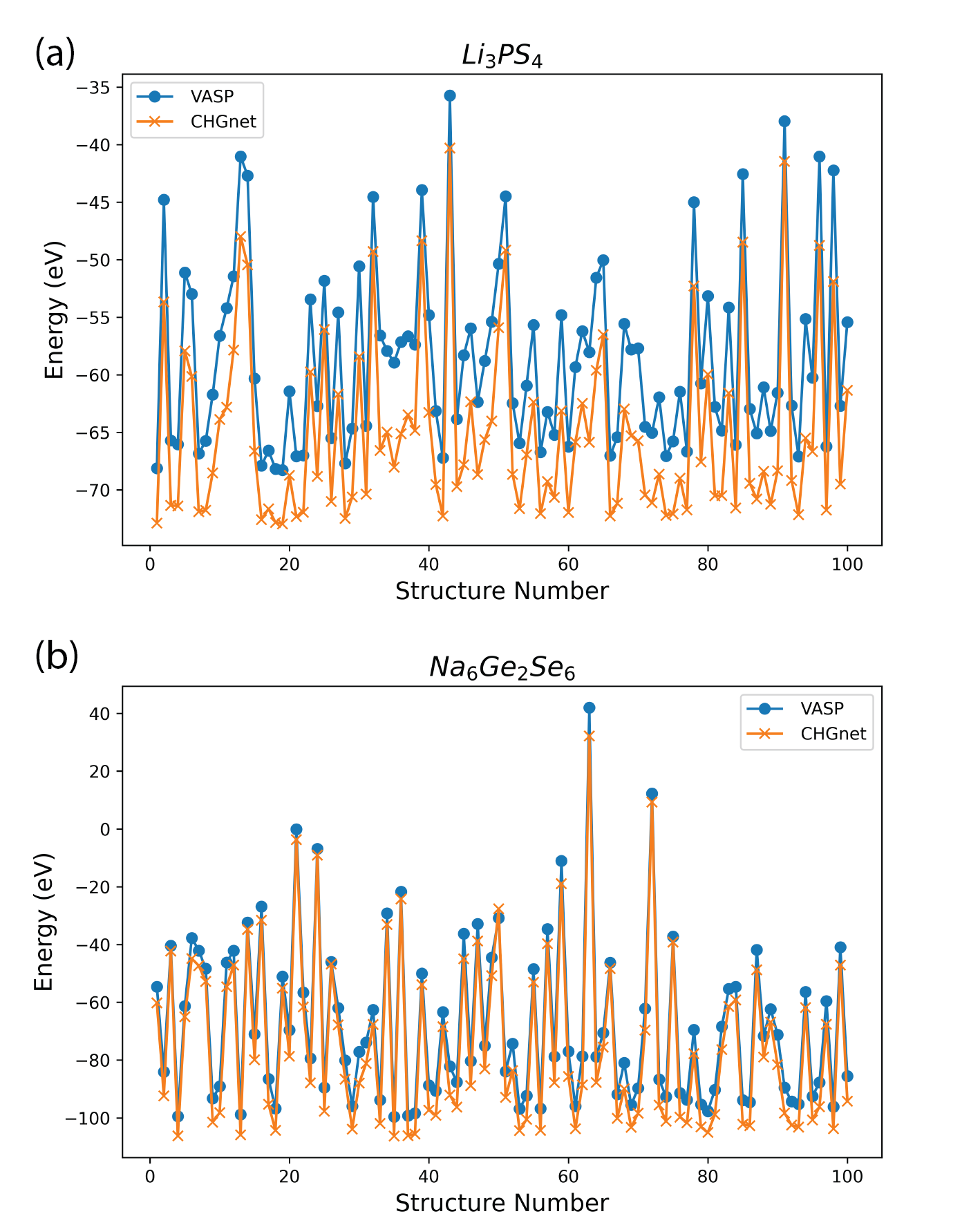}
    \caption{ \label{fig:ml}(a) The energy of randomly generated structures comparison between VASP and CHGnet for (a) \ch{Li3PS4} and (b) \ch{Na6Ge2Se6}}
\end{figure}

The maximum stoichiometric units for \ch{Li3PS4} and \ch{Na6Ge2Se6} are three and two respectively in our simulations. However, post-simulation symmetry analysis can exceed these limits. For example, the \ch{Li3PS4} phase in Figure \ref{fig:ps4_struc}(c) contains four stoichiometric units, yet it emerged as a low-energy structure from the CSP results in $P1$(1) with only two stoichiometric units. Using the Pymatgen symmetry analyzer, we found that the atoms of the relaxed structure with two \ch{Li3PS4} stoichiometric units occupy special WPs in a higher symmetry group with larger multiplicities. Similarly, the \ch{Na6Ge2Se6} structures in Figures \ref{fig:gese_struc}(a) and \ref{fig:gese_struc}(b) have three stoichiometric units, despite the simulation cell containing only two. Although we limited the number of atoms in our simulations, it is still possible to identify low-energy structures with a larger number of atoms.

Figure \ref{fig:energy_compare} illustrates the energy comparison between experimentally observed and predicted phases for \ch{Li3PS4} and \ch{Na6Ge2Se6}. For \ch{Li3PS4}, we include the calculated energies of the experimentally observed $\alpha$- and $\beta$-phases of \ch{Li3PS4} as reported\cite{kam2023crystal} at $0$ K. Notably, all candidates for metastable states possess higher energies than the the $\alpha$-, $\beta$-, and $\gamma$-phase, aligning with the fact that these are the only phases observed to date. The next lowest energy structure in space group $I\overbar{4}2m$(121) is merely 1 meV/atom higher than the $\alpha$-phase. For \ch{Na6Ge2Se6}, the predicted phase in $R\overline{3}$(148) has a significantly lower energy than the observed phase. Its dynamical stability suggests that this phase might be experimentally observable. Further exploration of other properties of this structure is warranted, although it lies outside the scope of this paper.

The proposed method successfully identified the experimentally observed phases for \ch{Li3PS4} and \ch{Na6Ge2Se6} along with several metastable candidates, demonstrating that focusing on the most promising regions of the phase space by fixing rigid blocks is an effective strategy for complex chalcogenides. There are other CSP tools that also support this feature, such as USPEX which utilizes evolutionary algorithms. One of the main differences between our methods is that we restrict atoms to move only within their designated WPs. Consequently, the CSP simulation is strictly confined to the pre-determined space group. As shown in Figure \ref{fig:ps4_csp}, this restriction significantly increases the chances of visiting the lowest energy phase in space group $Pmn2_1$(31) compared to $P1$(1), $P2_1$(4), and $Pc$(7), which have only one general WP without any restrictions on atomic positions. However, the WP restriction prohibits the situation where a lower energy phase requires a transition between WPs for atoms. For instance, Figure \ref{fig:gese_csp} illustrates that although the CSP simulation in space group $P2_1/c$(14) frequently visited the observed \ch{Na6Ge2Se6} phase, it was unable to identify even lower energy structures that were found in $P1$(1).

\begin{figure*}[t]%
    \includegraphics[width=18cm]{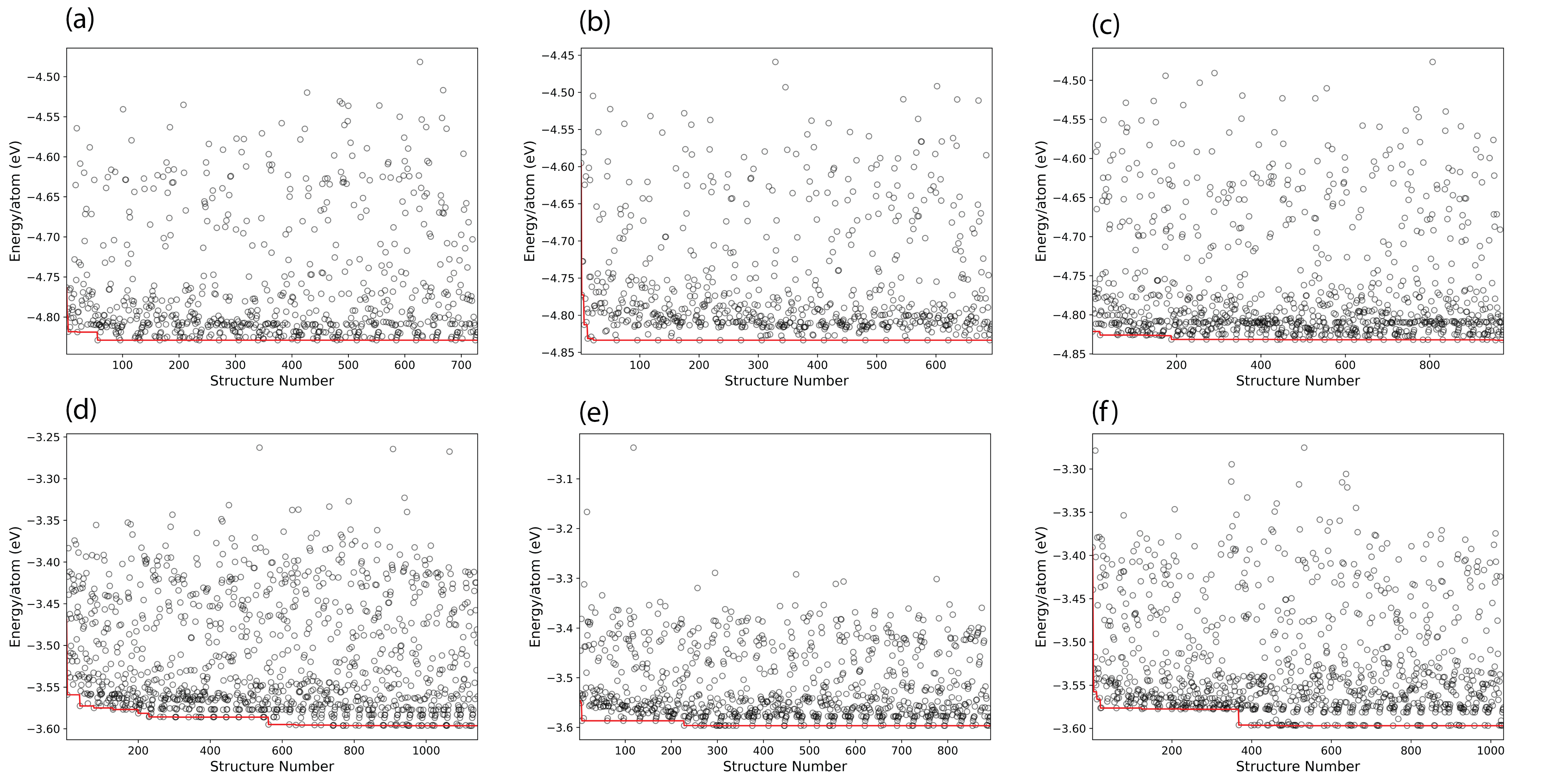}
    \caption{\label{fig:uspex}The USPEX CSP results for \ch{Li3PS4} and \ch{Na6Ge2Se6}. (a)(b)(c): \ch{Li3PS4}, (d)(e)(f): \ch{Na6Ge2Se6}}
\end{figure*}

While we impose a WP restriction in our method, USPEX employs mutation operations to deviate from the current state and explore more regions of the PES. This is one of the key aspects contributing to the performance of a CSP method. However, there may be limitations in identifying metastable phases compared to the WP restrictions we imposed in our method. We conducted the same simulations using USPEX for the \ch{Li3PS4} and \ch{Na6Ge2Se6} systems with fixed \ch{PS4} and \ch{Na6Ge2Se6} rigid blocks. The results are presented in Figure \ref{fig:uspex}. The USPEX simulations demonstrate a more systematic approach to evolving towards the lowest energy phase compared to the random sampling scheme we use. For \ch{Li3PS4}, USPEX found the same two lowest energy structures as we did, depicted in Figures \ref{fig:ps4_struc}(d) and \ref{fig:ps4_struc}(e). However, it found the structure in \(I\overbar{4}2m\) (121) with a higher energy in only one out of three trials. In contrast, our method identified this structure multiple times in the CSP results within the space groups $P1$(1), $P2_1$(4) and $Pc$(7), which are subgroups of $I\overbar{4}2m$(121). 
More importantly, for \ch{Na6Ge2Se6}, USPEX failed to find the experimentally observed phase in all three trials, despite all trials converging to the same lower energy structure illustrated in Figure \ref{fig:gese_struc}(b). In contrast, our method identified the observed phase not only in the CSP results within space group $P2_1/c$(14) but also in its subgroups $P2_1$(4) and $Pc$(7).

\begin{figure*}[t]%
    \includegraphics[width=18cm]{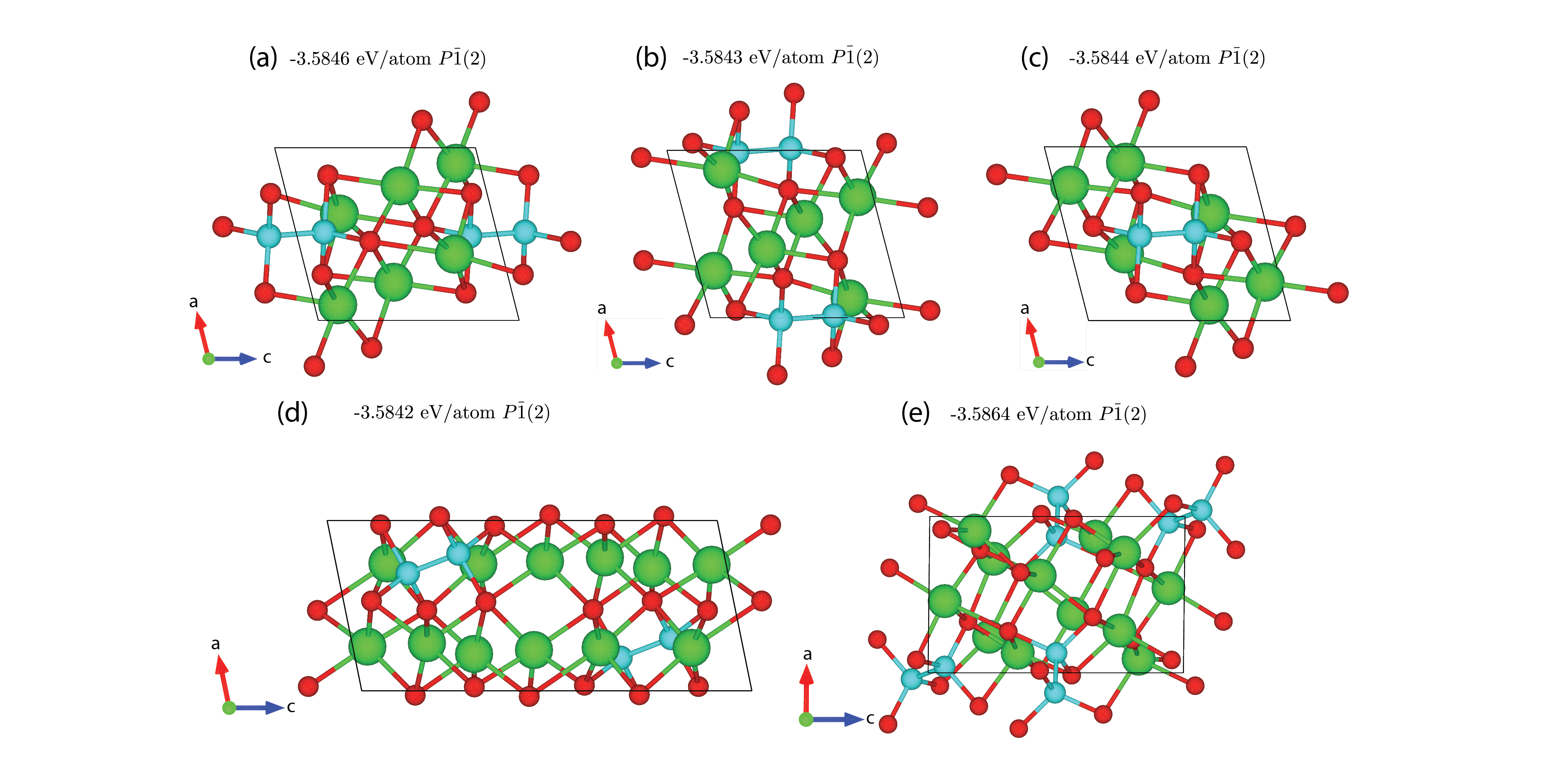}
    \caption{\label{fig:gese_sym2}The low energy structure found crystalized in space group $P\overbar{1}$(2)}
\end{figure*}

The possible reason for that is the \ch{Na6Ge2Se6} system features many structures crystallizing in space group $P\overbar{1}$(2) with lower energy than the observed phase (Figure \ref{fig:gese_sym2}). The success of evolutionary algorithms depends on their ability to inherit the best structural patterns from their parent structures. With so many lower energy structures distinct from the observed one, the likelihood of visiting the configurations of the observed phase on the PES is significantly reduced when relying mainly on random structure generations and mutations. On the other hand, since we imposed WP restrictions in our method, it is prohibited to visit those phases with lower symmetry in \(P\overbar{1}\)(2) while conducting CSP simulations in space groups $P2_1$(4), $Pc$(7) and $P2_1/c$(14). Consequently, this increases the likelihood of identifying the observed phase with higher energy.

We also evaluated the accuracy of MLP CHGnet\cite{deng2023chgnet} in predicting energy of structures. For unrelaxed structures, We randomly generated a hundred structures for \ch{Li3PS4} and \ch{Na6Ge2Se6}, then evaluated their energy using VASP and CHGnet. The results, shown in Figure \ref{fig:ml}, demonstrate the remarkable performance of CHGnet. Accurately determining the energy difference between two consecutive states is crucial for SA. CHGnet achieved 96\% and 100\% accuracy in predicting the lower energy state among two consecutive states for \ch{Li3PS4} and \ch{Na6Ge2Se6}, respectively. Regarding the absolute energy difference values, CHGnet managed to predict within a 16\% difference compared to VASP. For the \ch{Li3PS4} system, the percentage difference increased to 37\%. However, CHGnet's accuracy is not sufficient in differentiating the lower energy phase in relaxed structures. For \ch{Na6Ge2Se6}, it still maintains relatively high performance, successfully predicting the lowest energy phase. However, it predicted the energy of number three \ch{Na6Ge2Se6} structure in Figure \ref{fig:energy_compare}(b) to be 0.12 meV/atom lower than that of number four, whereas VASP results indicates it is 0.21 meV/atom higher. For \ch{Li3PS4}, CHGnet predicted a lower energy for the $I\overbar{4}2m$(121) phase than for the $\gamma$-phase by 14 meV/atom, contrary to the VASP results. The outstanding performance of CHGnet in predicting the energies of unrelaxed structures has led us to incorporate it into the latest implementation of our method as an option for calculating the energy of unrelaxed structures during the SA process.

The main drawback of the proposed method is that, even when limiting the space groups for CSP simulations, the total number of runs can still accumulate to a few hundred. Although they can be run concurrently as independent processes, making efficient use of an High-performance computing (HPC) environment, having too many individual simulations to conduct increases the likelihood of errors and raises the computational cost. However, we can mitigate this by focusing only on CSP in some low-symmetry space groups with a few WP combinations. In the results for the two systems illustrated in this paper, our method identified more metastable candidates in $P1$(1), $P2_1$(4) and $Pc$(7) than the results from USPEX, with fewer structures relaxed. Another limitation is that our method currently only supports CSP for determined chemical compositions. Implementing support for multinary systems without pre-determined compositions is currently being explored and will be added in the future.

\section{Conclusions}

We developed a new symmetry-restricted CSP method for structures with determined chemical compositions, specifically suited for predicting structures with rigid bodies. The validity and effectiveness of this method have been demonstrated with two metal chalcogenides, \ch{Li3PS4} and \ch{Na6Ge2Se6}. In both cases, we not only identified the experimentally observed phases but also uncovered several potential metastable states. Notably, for \ch{Na6Ge2Se6}, our method identified a dynamically stable phase with significantly lower energy, as evaluated by DFT calculations. This suggests that the experimentally observed phases might not be the ground state at absolute zero temperature. The results were compared with a popular CSP package, USPEX, and our method showed better performance in predicting metastable structures. The Python code to implement the proposed method can be found at \url{https://github.com/ColdSnaap/sgrcsp.git}


\begin{acknowledgments}
The authors thank NSF, USA (grant No. DMR1809128) for the funding of this project. The authors also acknowledge the usage of the HPC cluster "Foundry" at Missouri S\&T funded by NSF award OAC-1919789.
\end{acknowledgments}


\section*{data avaiability statement}
The data that support the findings of this study are available upon request.

\nocite{*}
\section*{References}
\bibliography{lib.bib}




\begin{table*}[t]
\centering
\caption{\label{tab:Li3ps4Energy}
Lowest energy/atom above the ground state found in different space groups, \ch{Li3PS4}}
\begin{ruledtabular}
\begin{tabular}{cccccc}
 Space group & Energy/Atom(meV) & Space group & Energy/Atom(meV) & Space group & Energy/Atom(meV)\\
\hline & \\[-1.em]
1 & 0 & 4 & 0 & 5 & 11\\
7 & 0 & 8 & 13 & 11 & 12\\
18 & 23 & 26 & 57 & 30 & 32\\
31 & 0 & 32 & 71 & 34 & 53\\
35 & 41 & 38 & 87 & 44 & 66\\
75 & 76 & 85 & 45 & 86 & 47\\
90 & 33 & 94 & 50 & 100 & 67 \\
101 & 86 & 102 & 43 & 114 & 54 \\
121 & 8 & 129 & 32 & 134 & 46\\
137 & 53 & 144 & 56 & 145 & 185 \\
146 & 62 & 151 & 137 & 152 & 70 \\
153 & 56 & 154 & 63 & 159 & 55 \\
160 & 146 & 171 & 216 & 172 & 156\\
173 & 117 & 185 & 93 & 186 & 82 \\
197 & 67 & 201 & 106 & 208 & 74 \\
215 & 145 & 217 & 104 & 218 & 96 \\
224 & 55 &  &  &  &

\end{tabular}
\end{ruledtabular}
\end{table*}

\begin{table*}[t]
\caption{\label{tab:GeSecases}
Numbers of combinations for \ch{Na6Ge2Se6} in different space groups with a maximum of two \ch{Ge2Se6} rigid bodies in the unit cell.}
\begin{ruledtabular}
\begin{tabular}{cccccc}
Space group & Combinations & Space group & Combinations & Space group & Combinations\\
\hline
\colorbox{blue!30}{1} & 2 & 2 & $>$1000 & 3 & $>$1000\\
\colorbox{blue!30}{4} & 1 & 5 & 32 & 6 & 228\\  
\colorbox{blue!30}{7} & 1 & \colorbox{blue!30}{8} & 4 & 10 & $>$1000\\
11 & 140 & 13 & 632 & \colorbox{blue!30}{14} & 16\\
16 & $>$1000 & 17 & 520 & 18 & 32\\
25 & $>$1000 & 26 & 32 & 27 & 520\\
28 & 150 & 30 & 32 & \colorbox{blue!30}{31} & 4\\
32 & 32 & 34 & 32 & 49 & $>$1000\\ 
51 & $>$1000 & 53 & 228 & 55 & 228\\ 
58 & 136 & 75 & 84 & 77 & 150\\
81 & 828 & 83 & 874 & 84 & 616\\
143 & $>$1000 & 147 & 151 & 149 & $>$1000\\
150 & 244 & 156 & $>$1000 & 157 & 62\\
158 & 117 & 159 & 24 & 162 & 247\\
163 & 120 & 164 & 247 & 165 & 22 \\
168 & 27 & 173 & 24 & 174 & $>$1000 \\
176 & 30 & 177 & 272 & 182 & 124\\ 
182 & 124 & 183 & 37 & \colorbox{blue!30}{185} & 8\\
186 & 26 & 187 & $>$1000 & 188 & 417 \\
189 & 213 & 190 & 31 & \colorbox{blue!30}{193} & 18\\
194 & 32
\end{tabular}
\end{ruledtabular}
\begin{minipage}{18cm}
\vspace{0.1cm}
\begin{flushleft}
\footnotesize Note: We have applied the proposed CSP method to the highlighted space groups, which have 20 or fewer combinations. Combinations exceeding 1000 are marked as '$>1000$'. Space groups not included in the table have no viable cases.
\end{flushleft}
\end{minipage}
\vspace{0.3cm}
\caption{\label{tab:GeSeEnergy}
Lowest energy/atom above the ground state found in different space groups, \ch{Na6Ge2Se6}}
\begin{ruledtabular}
\begin{tabular}{cccccc}
 Space group & Energy/Atom(meV) & Space group & Energy/Atom(meV) & Space group & Energy/Atom(meV)\\
\hline & \\[-1.em]
1 & 0 & 4 & 17 & 7 & 18\\
8 & 34 & 14 & 17 & 31 & 36\\
185 & 32 & 193 & 27 \\
\end{tabular}
\end{ruledtabular}
\end{table*}

\end{document}